\documentclass[preprint]{elsarticle}

\usepackage{amsmath,amssymb}

\newcommand{\f}[2]{\frac{#1}{#2}}
\newcommand{\Oc}{{\cal O}}

\newcommand{\nyp}[3]{{\bf #1} (#2) #3}
\newcommand{\jhep}{Jour.\ High Energy Phys.\ }

\begin{document}

\begin{frontmatter}

\title{Comments on high-energy total cross sections in QCD}

\author[l1]{Matteo Giordano}
\ead{giordano@atomki.mta.hu}
\author[l2]{Enrico Meggiolaro}
\ead{enrico.meggiolaro@df.unipi.it}
\address[l1]{Institute for Nuclear Research of the Hungarian Academy of
  Sciences (ATOMKI), Bem t\'er 18/c, H--4026 Debrecen, Hungary}
\address[l2]{Dipartimento di Fisica, Universit\`a di Pisa,
and INFN, Sezione di Pisa, Largo Pontecorvo 3, I--56127 Pisa, Italy}


\begin{abstract}
We discuss how hadronic total cross sections at high
energy depend on the details of QCD, namely on the number of colours
$N_c$ and the quark masses. We find that while a ``Froissart''-type
behaviour $\sigma_{\rm tot}\sim B\log^2s$ is rather general, relying
only on the presence of higher-spin stable particles in the spectrum,
the value of $B$ depends quite strongly on the quark masses. Moreover,
we argue that $B$ is of order $\Oc(N_c^0)$ at large $N_c$, and we
discuss a bound for $B$ which does not become singular in the $N_f=2$
chiral limit, unlike the Froissart-\L ukaszuk-Martin bound.
\end{abstract}



\end{frontmatter}
%


\section{Introduction}
\label{sec:intro}

The behaviour of hadronic total cross sections at high energy is one
of the oldest puzzles of strong interactions. Experimental results, up
to the largest energies available at hadronic
colliders~\cite{TOTEM1,TOTEM2,TOTEM3,TOTEM4}, show a steady rise of
total cross sections for $\sqrt{s}\gtrsim 5~{\rm GeV}$~\cite{pdg},
where $s$ is the total center-of-mass energy squared. The theoretical
challenge is to explain the observed behaviour starting from the
first principles of QCD, which is believed to be the fundamental
theory describing strong interactions. So far, most of the efforts 
have focussed on phenomenological approaches, aimed at finding the
appropriate parameterisation of experimental data, usually taking
inspiration from the Regge-Gribov theory. To date, the majority of the
parameterisations agree on the leading energy dependence being of the
``Froissart''-type $\sigma_{\rm tot}\sim B_{\rm exp}\log^2
s$ with universal $B_{\rm
  exp}$~\cite{pdg,COMPETE,II,bd1,HIIK,bd4,bd5}, although  
alternative behaviours are also considered~\cite{FMS,DL}. 
A universal $\log^2 s$ rise, first proposed by
Heisenberg~\cite{Heisenberg}, has been supported by several
theoretical
arguments~\cite{CW,soft-pomeron1,soft-pomeron2,BKYZ,CGC1,CGC2,DGN},  
and recently also by numerical results in lattice QCD~\cite{GMM}.  

A correct prediction (from first principles) of the high-energy
behaviour of total cross sections would nontrivially confirm
the validity of QCD as the fundamental description of strong
interactions, in a largely untested energy-momentum regime.
In fact, the main difficulty in attacking this problem in the
framework of QCD is its nonperturbative nature, as it is part of the
more general problem of {\it soft} high-energy scattering,
characterised by small transferred momentum squared   
$t$ ($|t| \lesssim 1\,{\rm GeV}^2$) and large $s$. 
To avoid the shortcomings of perturbation theory in the
presence of the soft scale $t$, a nonperturbative approach to these
processes has been 
developed~\cite{Nachtmann91,DFK,Nachtmann97,BN,Dosch,LLCM1,
pomeron-book,reggeon},   
which relates the relevant scattering amplitudes to the 
correlation functions of certain nonlocal operators, the so-called
Wilson loops, in the fundamental  theory. To our knowledge, this
approach is so far the closest to a systematic derivation from first
principles.

In a recent paper~\cite{sigtot} we have argued, within the above-mentioned
nonperturbative
approach~\cite{Nachtmann91,DFK,Nachtmann97,BN,Dosch,LLCM1,pomeron-book,reggeon}
in Euclidean space~\cite{analytic1,GM2009},    
that hadron-hadron total cross sections at high energy behave like
\begin{equation}
  \label{eq:cs}
  \sigma_{\rm tot} \sim B(1-\kappa)\log^2 s 
\le 2B\log^2s\,.
\end{equation}
The prefactor $B$ is determined from the stable asymptotic hadronic
spectrum, considering strong interactions {\it in isolation}, by
maximising the following ratio, 
\begin{equation}
  \label{eq:B}
  B = \max_{a,\,j^{(a)}>1} B^{(a)}\,,\qquad  B^{(a)} =\textstyle
  \left(\f{j^{(a)}-1}{M^{(a)}}\right)^2\,, 
\end{equation}
where $a$ runs over the particle species, and $j^{(a)}$ and $M^{(a)}$
are the spin and mass of particle $a$, respectively. Only higher-spin
particles ($j^{(a)}>1$) have to be considered: if they were
absent, then $\sigma_{\rm tot}$ would be at most a constant at high energy,
and $B=0$. The parameter $\kappa$ is bounded by unitarity to be
$|\kappa|\le 1$, but is otherwise undetermined at this stage. In
Ref.~\cite{sigtot} we remarked that the most natural value yielding
the universality observed in experiments is $\kappa=0$, corresponding
to a black-disk-like behaviour at high energy. However, we do not have
a purely theoretical argument to show that this is actually the case. 
Furthermore, the phenomenological analyses in the literature give
different estimates of the ``blackness'' of the scatterers in the
high-energy limit, see, e.g., Refs.~\cite{bd1,bd4,bd5,bd2,bd3}. 
In Ref.~\cite{sigtot} we also gave a numerical estimate of $B$ using 
experimental data for (QCD-)stable mesons, baryons and nuclear
states. The ``dominant'' particle, i.e., the one which maximises
$B^{(a)}$, turns out to be the $\Omega$ baryon, and yields $B_{\rm
  QCD}\simeq 0.56~{\rm GeV}^{-2}$, which compares well to the
experimental value $B_{\rm exp} \simeq 0.69\div 0.73~{\rm
  GeV}^{-2}$~\cite{pdg}. Interestingly enough, our value for $2B_{\rm
  QCD}$ is about two orders of magnitude smaller than the analogous
prefactor $B_{\rm FLM}=\f{\pi}{M_\pi^2}$ appearing in the 
Froissart-\L ukaszuk-Martin bound~\cite{FLM1,FLM2,FLM3}, and only
about $50\div 60$\% larger than the experimental value, resulting in a
much more restrictive ``Froissart-like'' bound (which is satisfied
by $B_{\rm exp}$). 

It is part of the standard lore that hadronic total cross sections
should be mostly governed by the ``gluonic sector'' of the theory, and
this leads to expect that they could be described fairly accurately
using the {\it quenched} approximation of QCD, i.e., pure $SU(3)$
gauge theory. In this case, and in the framework of the
nonperturbative approach discussed above, the relevant spectrum for
the computation of the prefactor $B$ would be the stable, higher-spin
part of the {\it glueball} spectrum. However, using data from
Ref.~\cite{MP}, the resulting value of $B$ turns out to be $2\div 3.5$
times the one obtained using the physical, {\it unquenched} spectrum,
suggesting the presence of unexpectedly large unquenching
effects~\cite{sigtot}. 

The high sensitivity of $B$ to the presence or not of dynamical 
quarks raises an interesting question: how much does the actual
value of $B$ depend on the details of QCD? More precisely, how much
does it depend on the values of its parameters, i.e., the number of
colours $N_c$ and the quark masses? Since only the stable spectrum
enters the maximisation Eq.~\eqref{eq:B}, the crucial point is to
understand how the stability of hadrons changes as the parameters are
varied, and how this affects the overall scale of total cross
sections. This is precisely the purpose of this paper. 
In section \ref{sec:largen} we discuss the large-$N_c$
limit. In section \ref{sec:chiral} we discuss the chiral limit. In
section \ref{sec:largem} we discuss the regime of large quark masses,
making contact with the quenched approximation. Finally, in section
\ref{sec:concl} we draw our conclusions.


\section{Large $N_c$}
\label{sec:largen}

We begin by discussing the behaviour of high-energy total cross
sections in the 't Hooft large-$N_c$
limit~\cite{'tHooft1,'tHooft2,Witten}.  
The first point we want to clarify is precisely how this limit has to
be taken. Eq.~\eqref{eq:cs} describes the asymptotic high-energy
behaviour of $\sigma_{\rm tot}$, i.e., $\sigma_{\rm tot}$ for
center-of-mass energies much larger than any other mass/energy scale
in the problem. Formally, Eq.~\eqref{eq:cs} has to be written as 
\begin{equation}
  \label{eq:cs2}
 \lim_{s\to\infty} \f{\sigma_{\rm tot}}{\log^2 s}= B(1-\kappa)\le 2B\,.
\end{equation}
The quantity $B$ is well defined for every finite $N_c$, as the number of 
stable states is finite, and so it is sensible to consider its large-$N_c$ 
limit.\footnote{Although exactly at $N_c=\infty$ there 
is an infinite tower of stable mesons with unbounded spin, so that 
$\max_a B^{(a)}$ may not exist there, this does not affect our limiting 
procedure.} 
It is therefore clear that we take {\it first} the large-$s$ limit,
and {\it then} the large-$N_c$ limit (differently, for example, from
what is done in Refs.~\cite{GdR,GdRV,Cohen}).
Taking, instead, {\it first} the large-$N_c$ limit, and {\it then}
the large-$s$ limit, the leading contribution to $\sigma_{\rm tot}$
comes from ``Pomeron exchange'' (understood here as the exchange
of gluons between the colliding mesons), and is of order $\Oc(1/N_c^2)$,
according to the usual counting rules.\footnote{The leading
contributions to meson-meson elastic scattering amplitudes are
actually expected to be of order $\Oc(1/N_c)$, and correspond to the
tree-level amplitudes of the large-$N_c$ mesonic effective
Lagrangian~\cite{Witten}. However, being real, forward tree-level
amplitudes do {\it not} give a contribution of order $\Oc(1/N_c)$ to
$\sigma_{\rm tot}$ via the optical theorem. A nonzero contribution of
order $\Oc(1/N_c^2)$ to $\sigma_{\rm tot}$ is obtained, instead, from
the imaginary part of the one-loop meson-meson elastic scattering
amplitudes. (We thank the referee for clarifying this point.)} 
In Ref.~\cite{Cohen} the two limits $s\to\infty$ and $N_c\to\infty$
are taken together, as the particles' momenta are scaled
proportionally to $\sqrt{N_c}$ as $N_c$ is increased. The resulting
total cross section is proportional to $\log^2 N_c$. In our approach
we do not have to scale the momenta, since they are formally taken to
infinity before taking the large-$N_c$ limit; all that matters is the
large-$N_c$ behaviour of the spectrum.  

The large-$N_c$ behaviour of meson and baryon masses is well
known~\cite{'tHooft1,'tHooft2,Witten}: meson masses are  
of order $\Oc(N_c^0)$, while baryon masses are of order
$\Oc(N_c)$. Roughly speaking, this is due to the fact that while
mesons are always $q\bar q$ states, independently of $N_c$, baryons
are made of $N_c$ quarks. Concerning higher-spin states, no
higher-spin QCD-stable meson is known in the ``real world'', i.e., for
$N_c=3$, and unless this is a subtle consequence of $\Oc(1/N_c)$
corrections to the meson masses at $N_c=\infty$, there is no reason to
expect the situation to change when $N_c$ is large (but finite). On
the other hand, a QCD-stable higher-spin baryon exists for $N_c=3$, namely
the $\Omega$ baryon ($j^{(\Omega)}=\f{3}{2}$). In the baryon sector,    
large-$N_c$ QCD possesses an effective light quark spin-flavour
contracted symmetry $SU(2N_f)$ for $N_f$ degenerate light quark
flavours~\cite{DJM1,DJM2}. Real-world QCD is close to have an exact
$N_f=2$ isospin symmetry, so for the physically most interesting case,
at large $N_c$ the contracted symmetry is $SU(4)$. Here we work with 
$2+1$ light flavours (up/down + strange), neglecting isospin breaking
effects. Furthermore, the large-$N_c$ limit is taken keeping $N_c$
odd, so that baryons are fermions as in the real, $N_c=3$ case.

Dashen, Jenkins and Manohar argued in Refs.~\cite{DJM1,DJM2} that in
terms of this emergent, large-$N_c$ symmetry, baryons can be
classified in multiplets corresponding to the irreducible
representations of the contracted spin-flavour symmetry. These
representations are labelled by the isospin $i$, the spin $j$, and a
further quantum number $k$, related to the number $N_s$ of strange
quarks as $N_s=2k$. The allowed values of $k$ for given $i,j$ are
obtained via the usual composition rule for angular momenta, so that
$|i-j|\le k \le i+j$.  
Large-$N_c$ consistency conditions, obtained by imposing unitarity in
pion-baryon and kaon-baryon scattering processes, constrain the form
of the baryon masses as follows~\cite{DJM1,DJM2},
\begin{equation}
  \label{eq:1}
  \begin{aligned}
    M &
= N_c m_0 + m_1 k + \f{1}{N_c}\left[ m_2 i(i+1)+ m_3 j(j+1)  + m_4
      k^2\right] + \Oc(1/N_c^2)\\ &  \equiv M_1(i,j,k) + \Oc(1/N_c^2)    
\,,
  \end{aligned}
\end{equation}
with mass parameters $m_i=m_i(N_c)$ which possess a $1/N_c$
expansion. 
This formula is valid for $j=\Oc(N_c^0)$, i.e., fixed spin as $N_c$ 
becomes large. 

The mass formula Eq.~\eqref{eq:1} is the starting point for the study
of the large-$N_c$ behaviour of the prefactor $B$ defined in Eq.~\eqref{eq:B}. 
Low-lying higher-spin states have masses differing from the lightest
baryon mass by terms of order $\Oc(j(j+1)/N_c)$, so for $j=\Oc(N_c^0)$
they will become stable at large enough $N_c$, since meson masses are
$\Oc(N_c^0)$ and so the available phase space for decays shrinks to
zero. The corresponding $B^{(a)}$ is of order $B^{(a)}=\Oc(1/N_c^2)$,
which leads to $\sigma_{\rm tot}$ behaving as expected according to
the na\"ive large-$N_c$ counting rules. However, it is also possible
that states with even higher spin are stable at large $N_c$, which
could change the large-$N_c$ behaviour of $B$. To see this, recall
that a state with a given value of $k$ is possible only if $k\le
i+j$. Furthermore, if $2k=N_s$ out of $N_c$ quarks are strange quarks,
one has $i\le N_c/2-k$, and so also $j\ge 2k-N_c/2$, which is
effective if $N_s\ge (N_c+1)/2$ (as $j\ge 1/2$). Consider now the
$\Omega$ baryon, defined for arbitrary $N_c$ as the baryon made of 
$N_c$ strange quarks, therefore having
$j^{(\Omega)}=k^{(\Omega)}=N_c/2$ and $i^{(\Omega)}=0$. In a
hypothetical decay of $\Omega$ into a baryon with $N_s\ge (N_c+1)/2$
strange quarks and spin $j$, one has from the bound above 
\begin{equation}
  \label{eq:2}
  \Delta j \equiv \f{N_c}{2}-j \le N_c -N_s \equiv 2\Delta k\,;
\end{equation}
in a hypothetical decay to a state with $N_s<(N_c+1)/2$, since
$2\Delta k > (N_c-1)/2$ and $\Delta j < (N_c-1)/2$, the bound
Eq.~\eqref{eq:2} still holds. As a consequence, a decay to a baryon
with a decrease of $\Delta j$ in spin has to come with at least
a decrease of $\Delta j$ in (the absolute value of) strangeness, which
requires the emission of $\Delta j$ kaons.\footnote{We keep assuming
  that no higher-spin meson becomes stable for $N_c<\infty$. Notice
  that decays into more baryons/antibaryons are forbidden at large $N_c$ by
  a negative mass difference of order $\Oc(N_c)$ between initial and
  final states.} It is therefore possible that the mass balance
between initial and final states remains negative, as it is
for $N_c=3$, therefore making the $\Omega$ stable also at large $N_c$.  

To make this statement quantitative one should know the exact mass
formula, rather than its approximation Eq.~\eqref{eq:1}, which in
principle is valid only for $j=\Oc(N_c^0)$. However, numerical studies
on the lattice~\cite{DeGrand} (up to $N_c=7$) find good agreement with
the mass formula Eq.~\eqref{eq:1} also for states with $j=\Oc(N_c)$. 
This indicates that higher-order terms in Eq.~\eqref{eq:1} give small
contributions even for $j=\Oc(N_c)$, so that they can be neglected 
(in a first approximation),
and Eq.~\eqref{eq:1} can be used to give a sensible quantitative
estimate of the stability of the $\Omega$ baryon at large $N_c$. 

Working in the isospin limit, one can estimate the mass parameters
$m_i$ at $N_c=3$ by fitting the (isospin averaged) masses of the
physical octet and decuplet baryons with the mass formula
Eq.~\eqref{eq:1}. The error on the masses is taken as the sum (in
quadrature) of the experimental error and of an extra uncertainty,
accounting for isospin breaking and electromagnetic effects. This
uncertainty is estimated as the standard deviation of the masses in an 
isomultiplet, and set to 1 MeV for isosinglets (raising this 
to $2\div 3$ MeV yields similar results). The fit of the baryon masses
with Eq.~\eqref{eq:1} yields effective parameters, which include
contributions from higher-order terms neglected in
Eq.~\eqref{eq:1}. To estimate the corresponding uncertainty 
$\varepsilon_{N_c}$, we have repeated the fit including an extra term
$\tilde m^{(a)}$ in the mass of each baryon, i.e., using the
expression $M^{(a)} = M_1(i^{(a)},j^{(a)},k^{(a)}) + \tilde m^{(a)}$
to fit the mass of baryon $a$. The parameters $\tilde m^{(a)}$ were
constrained to  be ``small'' by means of the usual constrained-fit 
techniques~\cite{bayes}. In particular, we took these extra parameters
to be normally distributed around zero with standard deviation
$\sigma=10\,{\rm MeV}$. This choice is motivated by the fact that they
are of order $\Oc(1/N_c^2)$, and that the simple fit indicates that
$m_i$ are of order $\Oc(100\,{\rm MeV})$. The results are reported in
Tab.~\ref{tab:2}. Variations of the resulting parameters between the
two fits give an estimate of $\varepsilon_{N_c}$, and are at most of
15\%. 

\begin{table}[t]
  \centering
  \begin{tabular}{ccccc}
 & $M_1$  &  $\varepsilon_{\rm stat}$ &  $M_1 +  \tilde m$ & $\varepsilon_{N_c}$ 
\\ \hline 
  $m_0$ &  $\phantom{-} 287.73$ &    $0.27$ &  $\phantom{-} 287.4$ & $0.3$\\
  $m_1$ &  $\phantom{-} 429.3$  &    $2.7$  &  $\phantom{-} 432$   & $3$  \\
  $m_2$ &  $\phantom{-} 101.8$  &    $2.5$  &  $\phantom{-} 97$    & $5$  \\
  $m_3$ &  $\phantom{-} 198.2$  &    $1.9$  &  $\phantom{-} 202$   & $4$  \\
  $m_4$ &  $-109.6$             &    $5.4$  &  $-125$   & $15$ \\
  $\delta m'$ & $-67$ & $10$ &  $-55$ & $12$
  \end{tabular}
  \caption{Results for the mass parameters $m_i$ and for $\delta m'$
    from the fit of the (isospin averaged) masses of the octet and
    decuplet baryons with the mass formula $M^{(a)} =
    M_1(i^{(a)},j^{(a)},k^{(a)})$, Eq.~\eqref{eq:1}  
    (first column), and with the formula $M^{(a)} =
    M_1(i^{(a)},j^{(a)},k^{(a)}) + \tilde m^{(a)}$, with constrained
    $\tilde m^{(a)}$ (third column). Statistical errors on the simple
    fit (second column) and uncertainties due to higher-order terms
    (fourth column) are also reported. Masses are in MeV.} 
  \label{tab:2}
\end{table}

All the parameters but $m_4$ are positive at $N_c=3$; if their sign
remains the same at large $N_c$ (which is supported by lattice
results~\cite{DeGrand}), it is easy to obtain a bound on the
mass balance $\Delta M_1$, plugging Eq.~\eqref{eq:2} into
Eq.~\eqref{eq:1},  
\begin{equation}
  \label{eq:3bis}
  \begin{aligned}
    \Delta M_1 &\equiv m_1\Delta k + \f{1}{N_c}\left[
      - m_2 i(i+1)
+ m_3 \Delta j\left(\f{N_c}{2}+j+1 \right) 
+ m_4 \Delta k \left(\f{N_c}{2}+k \right) 
\right] \\ &
\le \left[m_1 + 2m_3\left(1+\f{1}{N_c}\right)\right]\Delta k 
\mathop\to_{N_c\to\infty} 
\left[m_1 + 2m_3 \right]\!|_{N_c=\infty}\Delta k \,. 
  \end{aligned}
\end{equation}
Within our approximations, stability of the $\Omega$ baryon at large
$N_c$ is ensured if $(\Delta M_1-2\Delta k \cdot M_K) |_{N_c=\infty}< 0$
for all possible channels. Using the bound Eq.~\eqref{eq:3bis}, this
is certainly the case if $\delta m \equiv (m_1 + 2m_3
-2M_K)|_{N_c=\infty}< 0$. The numerical results of
Ref.~\cite{DeGrand} indicate that $|m_i|$ decreases as $N_c$ is
increased, so using $m_i(N_c=3)$ instead of $m_i(N_c=\infty)$ should
make the bound even more conservative. On the other hand, $\Oc(1/N_c)$
corrections to the meson masses have not been measured in lattice
simulations of the full theory. Numerical results for the quenched
theory~\cite{Bali} suggest that the variation of meson masses 
between $N_c=3$ and $N_c=\infty$ is of the order of 10\%. A reasonable
upper bound on $\delta m$ is therefore $\delta m \le m_1(3) + 2m_3(3)
- 2M_K(3)\cdot 0.9 \equiv \delta m'$. Our final result is
\begin{equation}
  \label{eq:final_estimate}
  \delta m' =  -67 \pm 10_{\rm stat} \pm 12_{N_c}\,{\rm MeV}\,. 
\end{equation}
We remind the reader that this bound is rather loose, since it does
not include the negative contribution of $m_4$, and it overestimates
$m_1$ and $m_3$. Moreover, $\delta m'$ remains negative up to a
reduction of around 15\% of $M_K$. 

Our conclusion is that stability of the $\Omega$ baryon at
large-$N_c$ is at least plausible. If it is indeed so, since the 
corresponding $B^{(\Omega)}$ is of order $B^{(\Omega)}=\Oc(N_c^0)$,
then one would necessarily have $B=\Oc(N_c^0)$. This is in contrast
with the expected $\Oc(1/N_c^2)$ from the na\"ive large-$N_c$ 
counting rules, but not in contradiction, as that expectation holds in
the limit $N_c\to\infty$ at large but fixed $s$.


\section{Chiral limit}
\label{sec:chiral}

We now turn to the chiral limit. More precisely, we consider the
$N_f=2$ chiral limit, with only the up and down quark masses set to
zero. In this case the spectrum of the theory contains three
massless pseudoscalar Goldstone bosons (the pions) due to the
spontaneous breaking of $SU(2)$ chiral symmetry. Note that massless
particles of spin 0 leave Eqs.~\eqref{eq:cs} and \eqref{eq:B}
unchanged~\cite{sigtot}.  

Generally speaking, the chiral limit can only turn stable states into
unstable states, and not viceversa, due to the possibility of decaying
through the emission of massless pions. This possibility however
does not concern the $\Omega$ baryon. Whether or not the $\Omega$
remains stable depends on how much its mass, and the masses of the
other strange baryons and of the kaon, change as the chiral limit is
approached. It is likely that the difference between the physical
masses of these particles and the corresponding masses in the 
chiral limit is of the order of the current light-quark masses, i.e.,
a few MeV. On the other hand, $M_\Omega - M_X - M_K \Delta N_s$ is
negative and of magnitude $\Oc(0.1 \div 1\,{\rm GeV})$ for all baryons
$X$ in the octet and in the decuplet, i.e., at least two orders of
magnitude larger than the expected effect of the chiral limit on the
kaon and strange baryons masses. The effect of this limit on the
masses of nuclei is again expected to be a few MeV, so we expect that
the $\Omega$ remains the dominant particle. An interesting consequence
of this result is that our ``Froissart-like'' bound,
Eqs.~\eqref{eq:cs} and \eqref{eq:cs2}, is not singular in the $N_f=2$
chiral limit. The Froissart-\L ukaszuk-Martin bound, on the other
hand, is singular in this limit since the prefactor $B_{\rm
  FLM}=\f{\pi}{M_\pi^2}$ diverges for massless pions.\footnote{ 
 The non-optimality of $B_{\rm FLM}$ had been already pointed out in
 Ref.~\cite{GdRV}.}
Concerning the double chiral/large-$N_c$ limit, it is likely that the
prefactor $B$ comes out to be of order ${\cal O}(N_c^0)$. Indeed, the
estimate given in section 2 is likely to remain valid due to the small
effect of having massless up and down quarks on the masses of
non-Goldstone particles.\footnote{ 
 The $N_f=3$ chiral limit (with also the strange quark mass set to
 zero) is more problematic: in this case also kaons become massless,
 and the $\Omega$ baryon  is no longer stable. The role of dominant
 particle will presumably be taken by  some stable, higher-spin
 nuclear state, but we cannot make any definite statement. In the
 double chiral/large-$N_c$ limit, as the masses of nuclei are likely
 to be of order ${\cal O}(N_c)$, one would still have $B={\cal
   O}(N_c^0)$ if there were  stable nuclei with spin ${\cal O}(N_c)$
 (but not larger), but again we cannot make  any definite statement.}


\section{Large quark masses}
\label{sec:largem}

Let us finally discuss the limit of large quark masses. For quark
masses larger than some critical value, purely gluonic states ({\it
  glueballs}) will become stable, and will enter the set over which
$B^{(a)}$ has to be maximised. Eventually, as the quark masses are
further increased, at most only a finite number of higher-spin mesons
and baryons will remain stable against decays, which can now take
place through the emission of glueballs, since these have finite
masses in the limit $m_q\to\infty$. Of course, the values of $B^{(a)}$
corresponding to mesons and baryons keep decreasing as the quark
masses increase. The bottom line is that for large enough quark
masses, the relevant part of the spectrum over which one has to
maximise $B^{(a)}$ will consist only of stable higher-spin
glueballs. Eventually, as $m_q\to\infty$, one will end up with the
quenched theory, where $B$ has been shown to be at least larger than
$B_Q\gtrsim 1.09~{\rm GeV}^{-2}$~\cite{sigtot}. 

It is interesting to remark that, according to our results, in the
problem at hand the full and the quenched theory are not equivalent in
the large-$N_c$ limit. This is essentially due to the fact that while
baryon masses grow like $N_c$, the stability of glueballs is not
improved as $N_c$ grows, since they can always decay into light mesons
(at least for physical quark masses), whose masses are essentially
unaffected by the large-$N_c$ limit. Therefore, glueballs do not enter
the game, while baryons still play an important role even though they
become heavier and heavier. It is however worth noticing that both the
quenched and the unquenched theory are expected to have $B=\Oc(N_c^0)$
at large $N_c$. Indeed, we have argued above that the full theory is
likely to show this behaviour due to the stability of the $N_c$-quark
$\Omega$ baryon. In the quenched theory, glueball masses are
$\Oc(N_c^0)$, and a few higher-spin stable states exist at
$N_c=\infty$ according to lattice results~\cite{LRR}, and so
$B_Q=\Oc(N_c^0)$.


\section{Conclusions}
\label{sec:concl}

In this paper we have discussed how hadronic total cross
sections at high energy depend on the details of QCD, namely on the
number of colours and the quark masses. The starting point 
is the relation between the overall scale of total cross sections and
the hadronic spectrum found in Ref.~\cite{sigtot}, in the
framework of the nonperturbative approach to soft high-energy
scattering~\cite{Nachtmann91,DFK,Nachtmann97,BN,Dosch,LLCM1,pomeron-book,
reggeon} in Euclidean space~\cite{analytic1,GM2009}. 

Our results indicate that
while a ``Froissart''-type behaviour $\sigma_{\rm tot}\sim B\log^2s$
is rather general, relying only on the presence of higher-spin stable
particles in the spectrum, the value of $B$ can depend quite strongly on
the details of the theory, and particularly on the quark masses. (For
example, it is likely to be discontinuous as the $N_f=3$ chiral limit 
or the limit $m_q\to\infty$ are approached.) On the other hand, we expect
that $B$ behaves smoothly as the large-$N_c$ or the $N_f=2$ chiral
limits are approached. There are three results that we want to
highlight in particular.  
\begin{itemize}
\item In the large-$N_c$ limit, $B$ is likely to be of order
  $\Oc(N_c^0)$, due to the stability of the $\Omega$ baryon,
  in contrast with the expectation based on the na\"ive counting
  rules. 
\item The more restrictive ``Froissart-like'' bound of
  Eqs.~\eqref{eq:cs} and \eqref{eq:cs2} is not singular in the $N_f=2$
  chiral limit, again due to the stability of the $\Omega$ baryon.
\item In the large-$N_c$ limit, the full and the {\it quenched} theory
  are not equivalent for what concerns total cross sections.
\end{itemize}

\section*{Acknowledgements}

MG is supported by the Hungarian Academy of Sciences under
  ``Lend\"ulet'' grant No. LP2011-011.

\end{document}